\documentclass{osa-article}
\usepackage{siunitx}
\usepackage{braket}
\usepackage{hyperref}

\journal{oe}
\articletype{Research Article}
\PassOptionsToPackage{normalem}{ulem}
\usepackage{ulem}
\providecolor{added}{rgb}{0,0,1}
\providecolor{deleted}{rgb}{1,0,0}

\begin{document}

\title{Asymmetric population of momentum distribution by quasi-periodically driving a triangular optical lattice}

\author{Xinxin Guo,\authormark{1} Wenjun Zhang,\authormark{1} Zhihan Li, \authormark{1} Hongmian Shui, \authormark{1} Xuzong Chen, \authormark{1} and Xiaoji Zhou\authormark{1,2,*}}

\address{\authormark{1}State Key Laboratory of Advanced Optical Communication System and Network, School of Electronics Engineering and Computer Science, Peking University, Beijing 100871, China\\
\authormark{2}Collaborative Innovation Center of Extreme Optics, Shanxi University, Taiyuan, Shanxi 030006, China}

\email{\authormark{*}xjzhou@pku.edu.cn} %% email address is required



\begin{abstract}
Ultracold atoms in periodical driven optical lattices enable us to investigate novel band structures and explore the topology of the bands. In this work, we investigate the impact of the ramping process of the driving signal and propose a simple but effective method to realize desired asymmetric population in momentum distribution by controlling the initial phase of the driving signal. A quasi-momentum oscillation along the shaking direction in the frame of reference co-moving with the lattice is formed, causing the formation of the mix of ground energy band and first excited band in laboratory frame, within the regime that the driving frequency is far less than the coupling frequency between ground band and higher energy bands. This method avoids the construction of intricate lattices or complex control sequence. With a triangular lattice, we experimentally investigate the influence of the initial phase, frequency, amplitude of the driving signal on the population difference, and observe good agreement with our theoretical model. This provides guidance on how to load a driving signal in driven optical lattice experiment and also potentially supplies a useful tool to form a qubit that can be used in quantum computation.
\end{abstract}

%%%%%%%%%%%%%%%%%%%%%%%%%%  body  %%%%%%%%%%%%%%%%%%%%%%%%%%
\section{Introduction}

Ultracold atom systems provide us unique opportunity to investigate the interaction between atoms and light\cite{cohen-tannoudji1998nobel,Zhang}, quantum many-body problems\cite{lewenstein2012ultracold,bloch2008manybody}, quantum precise metrology\cite{ludlow2015optical,Zhou}, etc. Since the experimental realization of Bose-Einstein Condensate (BEC)\cite{anderson1995observation,davis1995boseeinstein}, numerous works have been carried out to optimize the BEC\cite{Lu}, extract and eliminate noise pattern\cite{nlx,Cao} and build up atom interferometry\cite{rosi2014precision,Ramsey} and so on. Particularly, by forming certain interference pattern of laser beams, people could set up optical lattice (OL) systems, which provide us much insight into the condensed matter physics\cite{gross2017quantum,georgescu2014quantum,greiner2008condensedmatter,lewenstein2007ultracold}. In recent years, periodical driven triangular or honeycomb OLs received a great attention, because driven systems provide powerful tool to realize topological models\cite{cooper2019topological,arimondo2012chapter}, and triangular and honeycomb lattices are especially popular, due to the special symmetries existing in such systems\cite{struck2011quantum}, and thus they give the emerging of many novel band structures, quantum phases and dynamics\cite{messer2018floquet,flaschner2016experimental,zheng2014floquet}. For example, a paradigm topological model, the Haldane model\cite{haldane1988model}, has been experimentally realized\cite{jotzu2014experimental}, giving us much insight into the quantum Hall effect\cite{kane2005quantum,hsieh2008topological,chang2013experimental} and providing us the basis for exploring topological insulators and superconductors\cite{hasan2010colloquium}. Other interesting investigations regarding the topology include the Dirac points\cite{tarruell2012creating,wunsch2008diracpoint}, symmetry protected topological orders\cite{chen2012symmetryprotected}, frustrated magnetism\cite{struck2011quantum}, frustrated quantum anti-ferromagnetism\cite{eckardt2010frustrated} and so on.

To load atoms into a periodically driven system, one typically chooses to firstly adiabatically load the atoms into the first energy band of a static lattice, and then apply a sinusoidal driving function to the initial phase of the laser beams that are used to construct OLs in order to shake the lattice\cite{arimondo2012chapter,jotzu2014experimental,flaschner2016experimental,messer2018floquet}. However, this switching-on process of the driving signal could result in the heating of the interacting Bose-Einstein condensate in a driven optical lattice\cite{arimondo2012chapter} and thus the loss of physical information of the systems. To avoid this, people switch on the driving signal adiabatically, that is, linearly ramp up the amplitude of the driving signal (ramping up stage) and then hold the driving amplitude (holding stage). In this ramping up stage, the initial phase of the driving signal plays a significant role. Nevertheless, the impact of this stage on the initial state of the holding stage still need more investigation. 

In this work, we investigate the impact of the parameters of the ramping up stage on the initial state of the periodical driving stage, including the influence of the driving phase, amplitude and frequency. The driving signal is designed as a biased sinusoidal function with linearly ramped up amplitude, within the regime that the driving frequency is far less than the coupling frequency between the ground band and the excited bands. This driving signal results in an asymmetric population, where the population of the Bragg peaks in momentum distribution become asymmetric, while the positions of the Bragg peaks remain unchanged. In the frame of reference co-moving with the lattice, depending on the parameters (chosen value of  driving phase, amplitude and frequency) of the driving signal, the atoms experience an oscillation of quasi-momentum, and could be transferred to different desired quasi-momentum states. We also relate this asymmetric population to a coherent mix of s- and p- orbitals in the laboratory frame of reference. The s-orbitals make the majority of the population, and the p-to-s ratio could be extracted from the experimental time-of-flight (TOF) images. Here the initial phase $\varphi$ (ranging from 0 to $2\pi$) plays the critical role, as we will demonstrate below. Our theoretical model shows good agreement with experimental data. Since the theoretical model doesn't rely on the truth that the lattice is a triangular one, similar results are also applicable to other kinds of driven lattice.

The remainder of this manuscript is organized as follows. In Sec. \ref{sec:theoreticalModel}, we introduce the general theoretical model for realizing the orbital mixture. In Sec. \ref{sec:experimentImplementation}, we briefly introduce the implementation on our experimental triangular lattice and control sequence and give the theoretical results. The experimental results and analysis are given in Sec. \ref{sec:experimentResult}. Then we discuss the influence of other parameters and several potential application of this result in Sec. \ref{sec:discussion}, before the conclusions are given in Sec. \ref{sec:conclusion}. 

\section{Theoretical model for the asymmetric population}\label{sec:theoreticalModel}

\begin{figure}[htbp]
	\centering
	\includegraphics[width=\linewidth]{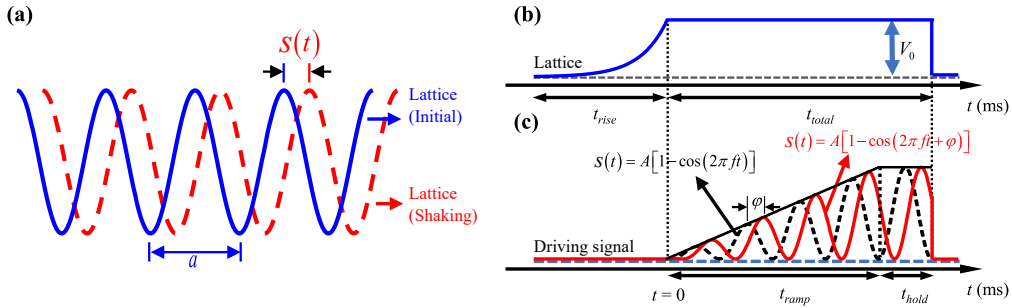}
	\caption{(a) The diagram for the driven lattice. The blue line denotes the initial position of the lattice. The red dashed line denotes the real-time position of the driven lattice. (b) shows the adiabatic loading time sequence, after which the atoms are populated onto the first band. $V_0$ represents the depth of optical lattice. (c) shows the time sequence of the driving signal $s(t)=	A\left[1-\cos\left(2\pi f t+\varphi\right)\right]$ , where $A$ is the driving amplitude, $f$ is the frequency and $\varphi$ is the initial phase of the shaking signal. The black dashed line shows the $\varphi=0$ condition, while the red solid line shows the non-zero $\varphi$ situation.}
    \label{fig:experimentalProcess}
\end{figure}

We first consider the general situation for the motion of atoms in a driven optical lattice. The optical lattice is constructed by a set of laser beams, whose intensity and initial phase determine the well depth and position of the lattice, respectively. The modulated variation of initial phase will thus result in the displacement of the lattice, whose time-dependent potential can be written as
\begin{align} 
	\label{eq:lattice potential}
	V\left(\vec{r},t\right) = -V_0\sum_{i<j}\cos\left[(\vec{k}_i - \vec{k}_j)\cdot\vec{r}+2\pi\cdot\frac{s_i(t) - s_j(t)}{a}\right],
\end{align}
where $k_i =  k =2\pi/\lambda $ is the wave number, $\lambda$ is the wavelength and $2\pi\cdot s_i(t) /a $ is the initial phase of the laser beam $i$, $V_0$ is the well depth, $a$ is the lattice constant and $s_i(t)$ denotes the displacement of the lattice with inital position $s_i(0)=0$. Fig. \ref{fig:experimentalProcess}(a) shows a simple one-dimensional example, for which we can assume $k_{i,j} = \pm x$. The blue (red) line denotes the lattice at its initial (real-time) position. Note the initial phase of a laser beam refers to the phase of its electric field at a specific space point (e.g., the emergent point of the beam from a mirror), distinguished with the phase of driving signal to be mentioned below.

In the preparation process, the lattice is adiabatically ramped up to $V_0$ (Fig. \ref{fig:experimentalProcess}(b)), where the atoms are transferred into the first band of the lattice. Immediately when the loading is completed, a sinusoidal function with linearly ramped up amplitude as shown in Fig. \ref{fig:experimentalProcess}(c), is applied as the quasi-periodical driving signal. 
\begin{equation}
	\label{eq:rampingsignal}
	s_i\left(t\right)=
	\begin{cases}
		A\dfrac{t}{t_{\text{ramp}}}\left[1-\cos\left(\omega t+\varphi\right)\right] & 0 < t < t_{\text{ramp}} \\
		A\left[1-\cos\left(\omega t+\varphi\right)\right] & t_{\text{ramp}} < t < t_{\text{total}}
	\end{cases},
\end{equation}
where $t_{\text{ramp}}$ is the ramping time, $t_{\text{hold}}$ is the holding time when the driving amplitude is fixed, $t_{\text{total}}=t_{\text{ramp}}+t_{\text{hold}}$ is the total time as shown in Fig. \ref{fig:experimentalProcess}(b) and \ref{fig:experimentalProcess}(c) ($t_{\text{total}}$, $t_{\text{ramp}}$ and $t_{\text{hold}}$ are all integer multiple of the sinusoidal function), $A$ is the driving amplitude, $\omega=2\pi f$ is the angular frequency and $\varphi$ is the initial phase of the shaking signal. Finally, the lattice and the driven signal are abruptly switched off at exactly the same time, before a time-of-flight (TOF) image is taken.

In laboratory frame of reference, the single-atom Hamiltonian can be written as 
\begin{align} 
\hat{H}_{\text{lab}}(\vec{r},t) = \frac{\hat{p}^2}{2m} + V(\vec{r},t).
\end{align}
The Schr\"{o}dinger equation for the center-of-mass is subjected to a unitary transformation to the frame of reference co-moving with the lattice, after which the Hamiltonian becomes
\begin{align} 
	\label{eq:co-movingHamiltonian}
	\hat{H}_{\text{co-moving}} = \frac{\hat{p}^2}{2m} -V_0\sum_{i<j}\cos\left[(\vec{k}_i - \vec{k}_j)\cdot\vec{r}\right] - \vec{F}(t)\cdot \vec{s}(t),
\end{align}
introducing a homogeneous inertial force $\vec{F}(t)$ acting in the co-moving frame, determined by the lattice motion according to
\begin{align}
    \label{eq:inertial force}
    \vec{F}(t) = -m \frac{d^2}{dt^2}\vec{s}(t).
\end{align}
Since the following discussions in this sub-section are all in the co-moving frame, for convenience, we leave out the subscript ``co-moving''. According to Bloch's theory, the motion of atoms in a optical lattice could be effectively described with its quasi-momentum $\hbar \vec{q}$ and effective mass $m^*$ in place of momentum $p$ and mass $m$, respectively. The motion equation reads
\begin{align}
    \label{eq:quasi-momentum-integral}
	\vec{q} = \frac{1}{\hbar}\int_0^{t_{\text{total}}}\vec{F}(t)dt = -\frac{m^*}{\hbar}\cdot \left.\frac{d}{dt}\vec{s}\right|_0^{t_{\text{total}}},
\end{align}
where $t_{\text{total}}$ is the total time as defined in Fig. \ref{fig:experimentalProcess}(b). Here we don't take the force of lattice into consideration, since the effect of lattice has been described as distortion of atoms' dispersion relation. A simple sinusoidal function lasts for integer periods won't work obviously, since the integral is zero. If the driving signal described by Eq. (\ref{eq:rampingsignal}) is along one specific axis, the total accumulated quasi-momentum along this axis is given by
\begin{align}
    \label{eq:quasi-momentum}
	q&=\frac{m^* \omega A}{\hbar}\sin\varphi \nonumber\\
	 &=Q\sin\varphi
\end{align}
where $Q\equiv{m^* \omega A}/{\hbar}$ is defined as the oscillation amplitude of quasi-momentum. 

By describing the motion of the atoms with its quasi-momentum and effective mass, one has actually assumed that the atoms will still occupy an energy eigenstate in the co-moving frame. Therefore, given the accumulated quasi-momentum, one can obtain the momentum distribution in the co-moving frame by simply solving the time-independent Schr\"{o}dinger equation, and then predict the momentum distribution in the lab frame by applying the inverse transformation. As presented in Sec. \ref{subsec:multi-orbital-mechanism}, this momentum distribution shows a  different population pattern, which characterises the formation of a mixture of s- and p- orbitals.

\section{Experimental setup and theoretical result}\label{sec:experimentImplementation}

\subsection{Experimental setup}

\begin{figure}[htbp]
	\centering
	\includegraphics[width=\linewidth]{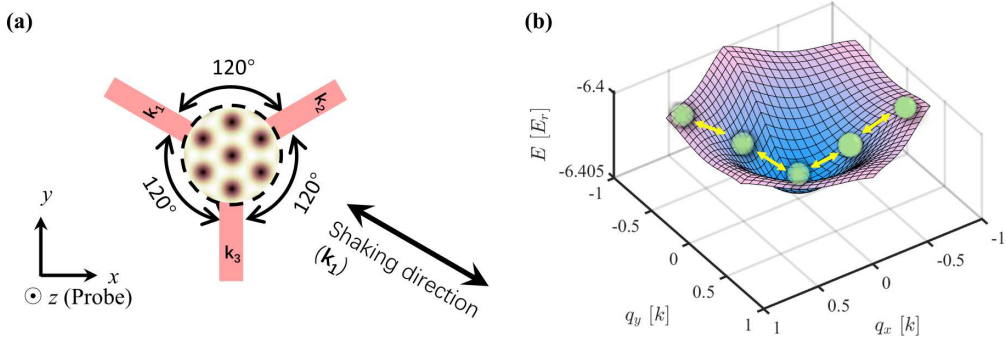}
	\caption{(a) The construction of the triangular lattice. Three linear perpendicular-to-plane polarized laser beam intersect at one point in $x$-$y$ plane, where the triangular lattice are formed. The three laser beams are denoted as $k_1$, $k_2$ and $k_3$. The initial phase of laser beam $k_1$ is controlled with a piezo-electric actuator. The driving signal is applied to this piezo-electric actuator to create shaking along direction $k_1$. The probe is set to perpendicular to the lattice plane (along axis $z$). (b) In the co-moving frame, this whole manipulation process can be understood as an oscillation of the quasi-momentum on the first band.}
    \label{fig:experimentalSetup}
\end{figure}

Our experimental demonstration is carried out in a 2D triangular lattice, where we could give a clearer  definition of the population difference. The experimental setup is similar to our previous work\cite{jin2018finite,xinhaozou}. The lattice potential is created by three intersecting \SI{1064}{\nano\metre} laser beams with \SI{120}{\degree} enclosing angle, which are linearly polarized perpendicular to the lattice plane. In preparation, a nearly pure condensate of about $1.5\times 10^5$ $^{87}$Rb atoms is obtained in a hybrid optical-magnetic trap with harmonic trapping frequencies $(\omega_x,\omega_y,\omega_z)=2\pi\times(28,55,65)\si{\hertz}$ in three directions respectively. Then the atoms are loaded adiabatically into the first band within $t_{\text{rise}}=\SI{60}{\milli\second}$. The driving signal is applied in the direction $\hat{k}_1$ as shown in Fig. \ref{fig:experimentalSetup}(a) with the help of a piezo-electric actuator. The probe is taken perpendicular to the lattice plane after \SI{30}{\milli\second} time-of-flight. 

\subsection{Theoretical result}

\begin{figure}[htbp]
	\centering
	\includegraphics[width=\linewidth]{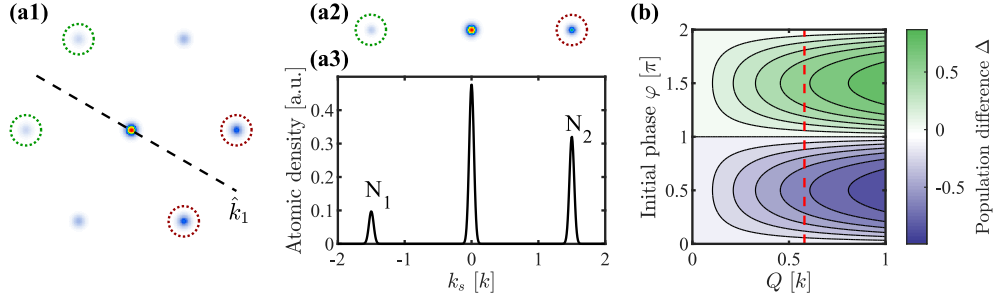}
	\caption{(a) Theoretical predicted momentum distribution of the atoms in the lab frame. (a1) shows the unbalanced momentum distribution, where the peaks in two red circles are sharper than those in two green circles. (a2) Side view of the momentum distribution along the direction $\hat{k}_1$ in (a1). (a3) Corresponding atomic density distribution in (a2). $N_1$ and $N_2$ denote the atom number of the left and right peak, respectively,  at $Q=k/6$ and $\varphi=\pi/2$. The notation $k_s$ denotes the quasi-momentum along the shaking direction k1. (b) Theoretical predicted population difference $\Delta$ with respect to the driving initial phase $\varphi$ and the quasi-momentum amplitude $Q$. When $Q$ is fixed (red dashed line), $\Delta$ experience periodic change as $\varphi$ increases. When $\varphi$ is fixed, $\Delta$ either rises or falls according to the chosen value of $\varphi$.}
    \label{fig:theoreticalPredict}
\end{figure}

As discussed before, in the co-moving frame the whole process can be well understood as an oscillation of quasi-momentum on the first band, as shown in Fig. \ref{fig:experimentalSetup}(b). In the co-moving frame, the quasi-momentum of atoms follows a certain path determined by our driving signal and periodically go around the designed path. According to Eq. (\ref{eq:quasi-momentum}), given the initial phase $\varphi$, driving amplitude $A$ and the frequency $f$, we can predict the momentum distribution in the lab frame as shown in Fig. \ref{fig:theoreticalPredict}(a1), by calculating the accumulated quasi-momentum of the atoms, obtaining the momentum distribution in the co-moving frame and then transforming it back to the lab frame. The atomic density peaks marked by the red dashed circles are sharper than that marked by the green dashed circles, i.e., the so-called population difference. We then take integral along the shaking direction $\hat{k_1}$ and get a one-dimensional side view image (Fig. \ref{fig:theoreticalPredict}(a2)), where the green (red) circles denote the addition of the two points in green (red) circles in Fig. \ref{fig:theoreticalPredict}(a1), and its corresponding atomic density distribution (Fig. \ref{fig:theoreticalPredict}(a3)). Clearly, the right peak (number of atoms being $N_2$) is higher than the left one (number of atoms being $N_1$). In order to quantify the population difference, we introduce $\Delta$ as 
\begin{align}
    \label{eq:def-difference}
    \Delta = \frac{N_1-N_2}{N_1+N_2}
\end{align}
where $N_1$ and $N_2$ are shown in Fig. \ref{fig:theoreticalPredict}(a3). 

The calculation of the relation between population difference $\Delta$ and quasi-momentum $q$ is done by numerically diagonalizing the Hamiltonian of an interaction-free atom in a lattice potential, which yields the eigen-energies and the eigen-states. Fig. 3(a) is an example plot of the theoretical eigen-state. The population involved in Fig. 3(a) and Fig. 3(b) are given by the inner-products between the eigen-state and corresponding momentum state.
Therefore, given the value of initial phase $\varphi$, amplitude $A$ and frequency $f$ of the shaking signal, we could give the numerical simulation of population difference $\Delta$ (defined in Eq. (\ref{eq:def-difference})) with respect to $\varphi$ and $Q$ (defined in Eq. (\ref{eq:quasi-momentum})) as shown in Fig. \ref{fig:theoreticalPredict}(b). When $Q$ is fixed (red dashed line), the population difference $\Delta$ will fall down first and then rise up as $\varphi$ increases from 0 to $2\pi$. When phase $\varphi$ is fixed, the population difference $\Delta$ will either rise or fall depending on whether the chosen value of $\varphi$ is larger or smaller than $\pi$. To achieve desired population difference or s-p mixture, one just needs to choose appropriate value of $A$ and $\varphi$ according to Fig. \ref{fig:theoreticalPredict}(b). Practically, $Q$ is fixed and one just needs to determine the value of $\varphi$. Our experimental data is obtained along this red dashed line.  

\subsection{Multi-orbital mechanism}\label{subsec:multi-orbital-mechanism}

\begin{figure}[htbp]
	\centering
	\includegraphics[width=\linewidth]{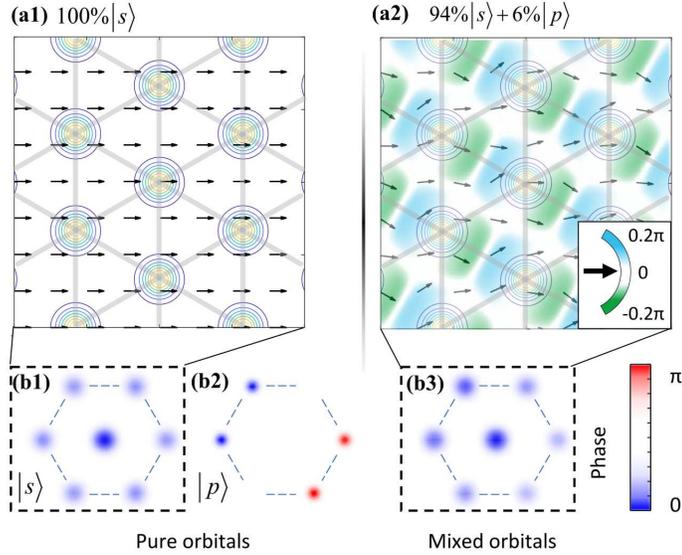}
	\caption{Multi-orbital mechanism of the population-difference state. The atomic distribution of pure s-orbital and the mixed s- and p- orbital in real space are shown in (a1) and (a2), respectively. The red and blue color or the arrows denote the local phase distribution. Momentum distribution of pure s-orbital (symmetric), pure P-orbital (anti-symmetric), and the mixed s- and p- orbitals are shown in (b1), (b2) and (b3), respectively. In (b3) the upper-left two points are enhanced and the lower-right two points are weakened.}
    \label{fig:spMix}
\end{figure}

In this manuscript we relate the unbalanced momentum distribution pattern shown in Fig. \ref{fig:theoreticalPredict} (a) to a coherent superposition of s- and p-orbitals. The mixed state is given by
\begin{align}
	\label{eq:superposition}
	\Psi = \sum_{n,q}c_{n,q}\cdot\psi_{n,q}
\end{align} 
where $\Psi$ is our observed state, $c_{n,q}$ is the superposition coefficients, $\psi_{n,q}$ is the Bloch state, $n$ and $q$ denotes the band index and quasi-momentum, respectively. In the shallow lattice regime, it's the first two bands $\vert s \rangle$ and$\vert p \rangle$ that play the dominant role. In Fig. \ref{fig:spMix}(a), (a1) shows the real space atom distribution of pure $\ket{s}$ state and (a2) the real space atom distribution of a mixed state of $94\%\ket{s}+6\%\ket{p}$ which is exactly the case in our experiment at $\varphi=0.75\pi$. The arrows and the red or blue color denote the phase of the wave function. In reciprocal space, Fig. \ref{fig:spMix}(b) provides a clear image of this mixed state. Because the wave function $\ket{p}$ is anti-symmetric, the upper left two points are enhanced, while the lower right two points are weaken. Therefore, an unbalanced picture is formed when the mixed state emerges (Fig. \ref{fig:spMix}(b2)). In addition, the proportion of $\ket{s}$ and $\ket{p}$ state varies as $\varphi$ changes, as shown in Fig. \ref{fig:experimentalFit}(a).

\section{Experimental demonstration}\label{sec:experimentResult}

In the experiment, the driven signal is chosen as $t_{\text{ramp}}=\SI{20}{\milli\second}$ and $t_{\text{hold}}=\SI{5}{\milli\second}$. We first investigated the influence of the initial phase $\varphi$ of the shaking signal, with frequency and amplitude fixed as $f=\SI{3}{\kilo\hertz}$ and $A=\SI{133}{\nano\metre}$ ($Q = 0.58k$) and well depth fixed as $V_0 = 4.0E_r$, where $E_r=(\hbar k)^2/(2m)$ is the recoil energy of an atom, to realize good visibility. In this situation, the band gap between the first and the second bands is around \SI{12}{\kilo\hertz}, which is much larger than our driving frequency. With various initial phase $\varphi$ (as the red dashed line shown in Fig. \ref{fig:theoreticalPredict}(b), we take TOF images of the atoms with time-of-flight being $t_{\text{TOF}} = \SI{31}{\milli\second}$. 

\begin{figure}[htbp] 
	\centering
	\includegraphics[width=\linewidth]{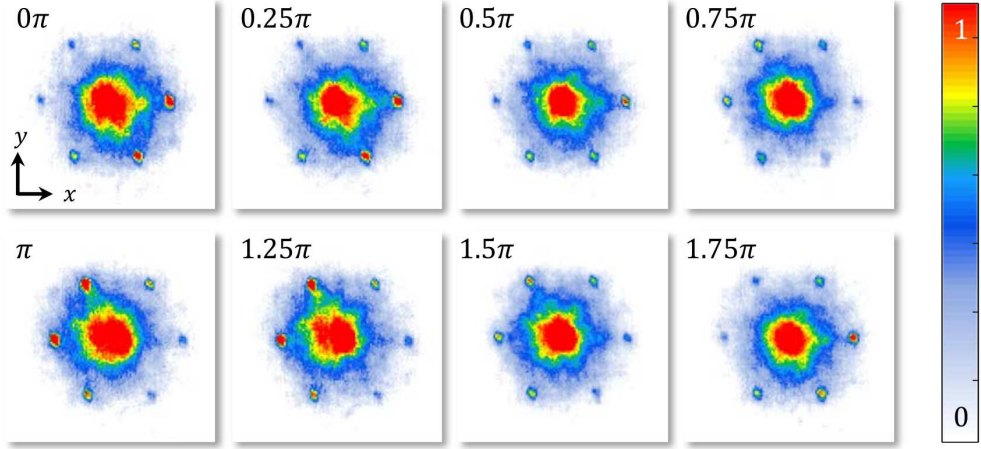}
	\caption{Observation of asymmetric momentum distribution at different driving initial phase $\varphi$ in experiment. As $\varphi$ increases, the momentum distribution shows a periodic change.}
	\label{fig:experimentalResult}
\end{figure}

In Fig. \ref{fig:experimentalResult} the experimentally obtained momentum distributions are shown for $\varphi$ ranging from \numrange{0}{2\pi} with step of $0.25\pi$ (each image is an average of five times of TOF images at the same condition). The $\varphi=0$ case isn't necessarily symmetric due to the imperfections in our experiment --- the shaking signal can't be switched off immediately, but with a little delay $\delta t$. As a result, the integration time in Eq. (\ref{eq:quasi-momentum-integral}) becomes $t_{\text{total}}+\delta t$. In addition, due to the imperfection in our phase control system, atoms accumulated an extra quasi-momentum in the shaking process, presented as a bias of the population difference shown in Fig. \ref{fig:experimentalResult}. The momentum distribution of atoms shows a deviation to the upper left at phase being larger than $0.5\pi$, and a deviation to the lower right at phase being 0 and $0.25\pi$. Meanwhile, the atoms experience a sudden ``kick'' at the very beginning of the driving process, resulting in a bias on the accumulated quasi-momentum. The above imperfections can be formulated in
\begin{align}
	\label{eq:true-quasi-momentum}
	q=Q\sin\varphi+\omega^2 A \delta t\cos\varphi + q_0
\end{align}
Taking this bias into consideration, this whole image shows a periodic change, allowing us to manipulate the population difference, or the quasi-momentum in the lattice frame of reference. 

\begin{figure}[htbp]
	\centering
	\includegraphics[width=\linewidth]{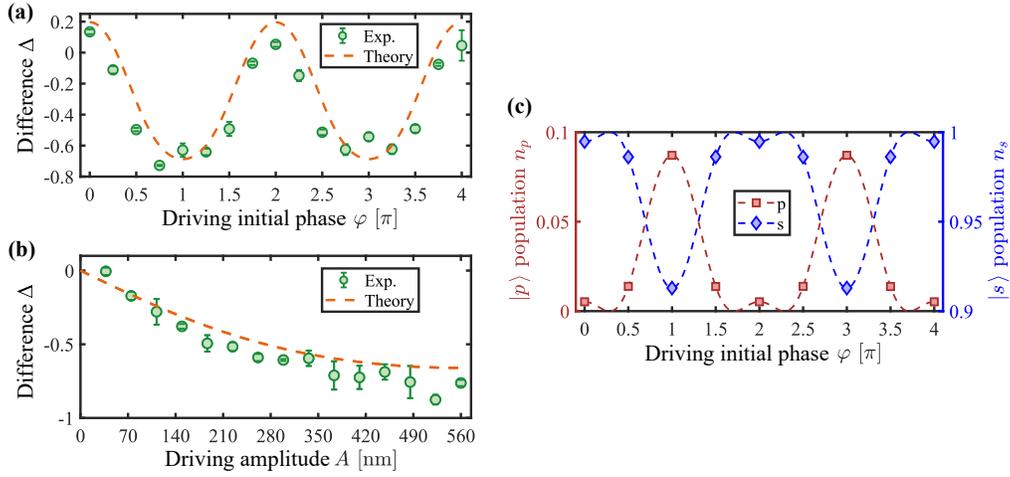}
	\caption{Experimental results. The green circles denote the experimental measurements, and the orange dashed line denotes the theoretical prediction. Each point shows an average of five independent measurements, and the error bar is given as the standard deviation during the averaging process. (a) The result of difference with respect to shaking initial phase $\varphi$. (b) With the initial phase fixed, the result of population difference changing with driving amplitude $A$. (c) The fraction of s- (blue) and p- orbital (red) corresponding to the same condition shown in (a). The square and diamond show the band fraction calculated from experimental data, while the dashed lines are the theoretical prediction. The p-orbital proportion is desired to be between 0 and 0.1, and the s-orbital to be between 0.9 and 1.}
    \label{fig:experimentalFit}
\end{figure}

The corresponding population difference is shown in Fig. \ref{fig:experimentalFit}(a), where the data is fitted with Eq. (\ref{eq:true-quasi-momentum}). As we can see from Fig. \ref{fig:experimentalFit}(a), the experimental results coincide well with the theoretical model, which indicates that this method is pretty precise and robust. 

The p-to-s ratio can be extracted from the population difference, as shown in Fig. \ref{fig:experimentalFit}(c). Given a lattice well depth, $\Delta$ is a one-to-one mapping function of the p-to-s ratio. Therefore, the p-to-s ratio can be calculated from $\Delta$. 

We also studied the effect of the amplitude of the shaking signal. With fixed initial phase $\varphi=0.8\pi$, the population difference $\Delta$ is shown in Fig. \ref{fig:experimentalFit}(b) for shaking amplitude of lattice ranging from \SI{35}{\nano\metre} to \SI{525}{\nano\metre}. When the shaking amplitude is small, the shaking doesn't contribute much to the momentum distribution of the atoms. As the shaking amplitude increases, the atoms experience more and more significant influence and the population difference becomes larger and larger. Eventually the atoms reach the state that the upper left two points totally disappear as is shown in Fig. 5.

\section{Discussion}\label{sec:discussion}

\begin{figure}
	\centering
	\includegraphics[width=0.6\linewidth]{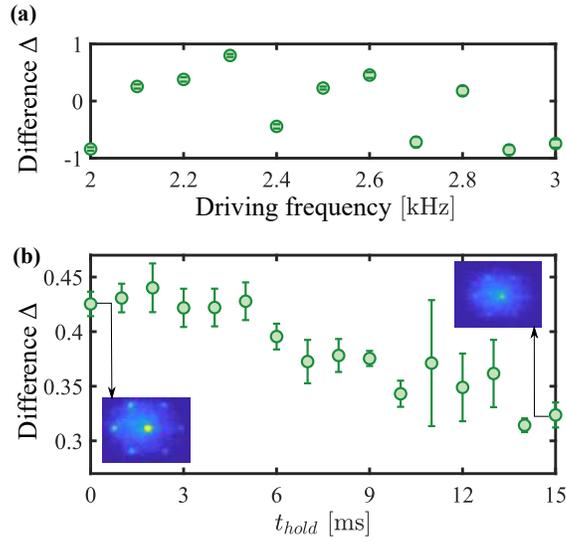} 
	\caption{(a) Population difference $\Delta$ versus driving frequency $f$. Measured at $V_0=4.0E_r$, $\varphi=\pi$, $A=\SI{266}{\nano\metre}$(b) Population difference $\Delta$ versus the holding time. Measured at $V_0=4.0E_r$, $\varphi=\pi$, $f=\SI{2.0}{\kilo\hertz}$. The inner plots show the corresponding TOF images of $t_{\text{hold}}=\SI{0}{\milli\second}$ (lower left) and $t_{\text{hold}}=\SI{15}{\milli\second}$ (upper right).}
	\label{fig:disscusion}
\end{figure}

As previously demonstrated, this method of realizing s-p orbital mixture by driving lattice can be perfectly applied to the triangular lattice, in the regime of appropriate shaking frequency $f$ and hold time $t_{\text{hold}}$. The chosen value of $f$ and $t_{\text{hold}}$ do have impact on the perfectness of this method, which are discussed as follows.

\paragraph{Effect of frequency} Although the experimental results are well demonstrated by our theoretical model, there does exist some discord in the influence of shaking frequency. As shown in Fig. \ref{fig:disscusion}(a), the relation between difference and frequency seems irregular. In this regime, our theoretical model is no longer suitable. In the derivation we directly neglect the force originated from the lattice but consider only the inertial force, which is only reasonable when the frequency is much larger than the harmonic frequency of the lattice potential well, and, at the same time, small enough compared to the corresponding frequency of the energy difference between the first band and higher energy bands. Therefore in other frequencies, either the lattice is too powerful, preventing the atoms from move in response to shaking, or the atoms could be excited to higher energy bands, where more complicated theoretical model is needed to consider.

\paragraph{Stability in long time} In the above investigation, the hold time of the shaking signal are chosen as \SI{5}{\milli\second} to achieve good visibility of experiment results. We would like to emphasize that s-p orbital mixture created by this method is quite stable. As shown in Fig. \ref{fig:disscusion}(b), when changing hold time from \numrange{0}{5} \si{\milli\second}, the population difference $\Delta$ remains a relatively constant value. Only after \SI{5}{\milli\second} can we observe unexpectedly shift in population difference. This shift is probably due to the de-coherence. In the regime of $t_{\text{hold}}>\SI{5}{\milli\second}$, the atom gas has actually shown indications of de-coherence and become heated due to the imperfections of other parts of our experiment system.

\section{Conclusion}\label{sec:conclusion}

In summary, we investigated the influence of the parameters of the ramping up stage of the driving signal on the initial state of the Floquet system, especially the initial driving phase. Basically, this driving signal with linearly ramping up amplitude results in an asymmetric population. We give our theoretical model of quasi-momentum-oscillation in the co-moving frame and experimentally demonstrate the asymmetric population. We also relate this asymmetric population to s-p orbital mixture of the a BEC in the optical triangular lattice. By shaking, the first band and the second band could be correlated together and the atoms could be transferred to different quasi-momentum state in the co-moving frame. The initial phase $\varphi$ of the driving signal could be chosen to achieve the designed population difference, or the designed quasi-momentum in the co-moving frame. The influence of driving amplitude are also investigated.  Furthermore, we point out this method is also useful in the cases of two-direction driving and other kind of lattices such as honeycomb lattice and square lattices. The experimental and theoretical results are well consistent with each other. This efficient method provides a probability to manipulate the population difference of the momentum distribution and the mix of different energy bands.

\section*{Funding}
National Program on Key Basic Research Project of China (No. 2016YFA0301501); National Natural Science Foundation of China (No. 61727819, No. 11934002, and No. 91736208).

\section*{Acknowledgments}

The authors would like to thank Xiaopeng Li for useful discussions.

%%%%%%%%%%%%%%%%%%%%%%% References %%%%%%%%%%%%%%%%%%%%%%%%%

\bibliography{pub-shaking-lattice}

\end{document}